\documentclass[11pt,oneside]{article}
\usepackage{authblk}
\usepackage[a4paper, top=1.0in, right =0.7in]{geometry}
\usepackage{graphicx}
\usepackage{psfrag}
\usepackage{amsmath}
\usepackage{amssymb}
\usepackage{amsfonts}
\usepackage{color}
\usepackage{natbib}
\usepackage{enumerate}
\usepackage{tikz}

\usepackage{comment}
\usepackage{pstricks}
\usepackage{epstopdf}
\usepackage{subfigure}
\usepackage{mathrsfs}
\usepackage{float}
\usepackage [english]{babel}
\usepackage{caption}
\usepackage{subcaption}
\usepackage{tikz}
\usetikzlibrary{shapes.geometric, arrows, positioning, calc}
\usepackage [autostyle, english = american]{csquotes}
\MakeOuterQuote{"}

\usepackage[left, pagewise]{lineno}

\newcommand{\bc}{\begin{center}}
\newcommand{\ec}{\end{center}}

\newcommand{\be}{\begin{equation}}
\newcommand{\ee}{\end{equation}}

\newcommand{\bea}{\begin{eqnarray}}
\newcommand{\eea}{\end{eqnarray}}
\newcommand{\beas}{\begin{eqnarray*}}
\newcommand{\eeas}{\end{eqnarray*}}

\newcommand{\bipin}[1]{\textcolor{blue}{#1}}

\newcommand{\tikzcircle}[2][red,fill=red]{\tikz[baseline=-0.5ex]\draw[#1,radius=#2] (0,0) circle ;}%
\DeclareRobustCommand\full  {\tikz[baseline=-0.6ex]\draw[thick] (0,0)--(0.5,0);}

\DeclareRobustCommand\dashed{\tikz[baseline=-0.6ex]\draw[thick,dashed] (0,0)--(0.54,0);}

\def\bx{{\bf x}}
\def\bX{{\bf X}}

\def\bu{{\bf u}}

\def\avg#1{{\left<{#1}\right>}}
\def\filt#1{{\widetilde{#1}}}

\title{\bipin{\underline{Preprint} } \\ Machine Learning-Based Estimation of Superdroplet Growth Rates Using DNS Data}
\author[1]{Divyaprakash}
\author[2]{Nikita N. Makwana}
\author[1]{Amitabh Bhattacharya}
\author[2]{Bipin Kumar\thanks{Corresponding author: bipink@tropmet.res.in}}
\affil[1]{Department of Applied Mechanics, I.I.T. Delhi, Hauz Khas, New Delhi  110016, India}
\affil[2]{Indian Institute of Tropical Meteorology, Ministry of Earth Science, Gov. of India, Pune, 411008 India}
\date{}

\begin{document}
\maketitle

\begin{abstract}
Droplet growth and size spectra  play a crucial role in the microphysics of atmospheric clouds. However, it is challenging to represent droplet growth rate accurately in cloud-resolving models such as Large Eddy Simulations (LESs). The assumption of "well-mixed" condition within each grid cell, often made by traditional LES solvers, typically falls short near the edges of clouds, where sharp gradients in water vapor supersaturation occur. This under-resolution of supersaturation gradients can lead to significant errors in prediction of droplet growth rate, which in turn affects the prediction of buoyancy at cloud edges, as well as forecast of precipitation. In "superdroplet" based LES model, a Lagrangian coarse-graining approach groups multiple droplets into superdroplets, each encompassing a specific number and size of actual droplets. The superdroplets are advected by the underlying LES velocity field, and the growth rate of these superdroplets is based on the filtered supersaturation field represented by the LES. To overcome the limitations of the "well-mixed" assumption, we propose a parameterization for superdroplet growth using high-fidelity Direct Numerical Simulation (DNS) data. We introduce a novel clustering algorithm to map droplets in DNS fields to superdroplets. The effective supersaturation at each superdroplet location is computed by averaging the unfiltered supersaturation of the associated droplets, which may differ from the value of filtered supersaturation at the superdroplet location. We then develop a machine learning-based parameterization to relate the effective growth rate of superdroplets to other filtered DNS flow variables. Preliminary results show a promising $R^2$ value of nearly 0.9 between the predicted and true effective supersaturation values for the superdroplets, for a range of superdroplet multiplicities.
\end{abstract}

\section{Introduction}

The dynamics of moist convection inside atmospheric clouds involves phenomenon occurring over a wide range of length scales \citep{grabowski2013growth}, ranging from microns (the size of typical cloud condensation nucleii) to kilometers (e.g. diameter of cloud base). Numerical simulation of atmospheric clouds can therefore be quite challenging, and resolving the condensation of individual droplets inside the clouds via Direct Numerical Simulations (DNSs) is not computationally feasible. Instead, Large eddy simulation (LES) based techniques \citep{moeng1986large,siebesma2003large} are typically used to numerically simulate atmospheric clouds \cite{siebesma2003large}. In LES, a spatial filter width $\Delta$ is used to demarcate the large inertial scale eddies from the smaller subgrid scale eddies. The large ``filtered" flow field is evolved exactly, and the effect of the subgrid scale eddies on the evolution of filtered flow field is modeled. LES of moist convection inside clouds involves coarse-graining the dynamics of droplet growth, for which two approaches can be taken. The Eulerian approach represents the droplets as a number density distribution function within each computational grid cell; this is also termed as the ``spectral" or ``bin" approach \citep{khain2004simulation}. The Lagrangian approach involves modeling the evolution of ``superdroplets" \citep{shima2009super}, which represent a cluster of $N_m$ actual droplets with similar attributes (e.g. size, velocity, mass of solute etc.), and is known to be more computationally efficient compared to the bin approach for modeling droplet collision, especially when the number of attributes increase. Both the bin approach and the Lagrangian superdroplet approach assume that the vapor field is well mixed in each grid cell, which in turn may lead to inaccuracies in prediction of droplet growth rates. This issue is particularly serious near the edges of clouds, where the gradients in the vapor supersaturation can be large.

In this work, we focus our efforts on improving the prediction of droplet growth rate in the superdroplet model \cite{shima2009super}, since it is known to be more computationally efficient compared to the bin based methods. Towards addressing the limitations of the well-mixed condition in LES of moist convection, we use  high-fidelity DNS data for entrainment and mixing of clear and cloudy air.
 Since last decade machine learning (ML) approach became popular in almost every field of science. In this regard, meteorology, in particular cloud microphysics, has been no exception. ML algorithms have been applied to DNS data for investigating the low and high vorticity regions \citet{Kumar2021, Nivelkar2023,Nivelkar2024}.  \cite{Frey2021} use DNS data and ML algorithms to improve turbulence modelling. A physics-informed ML approach was employed in LES in the study by \cite{Maejima2024}. 

 In this study, DNS data is used to train an ML model that predicts growth rate of the superdroplets in the transient filtered flow field. The ML model is trained for initial 5 seconds time duration, which is comparable to the initial evaporation time scale of the droplets. Superdroplets with a spatial distribution similar to the actual droplets are generated for each time instant. 
The actual droplets are mapped to superdroplets via a nearest-neighbor algorithm, and the effective growth rate of the superdroplet is calculated by enforcing mass conservation of the liquid in each superdroplet. An instantaneous effective supersaturation ($S_{eff}$) can then be calculated for each superdroplet, which will in general be different from the value of filtered supersaturation ($\filt{S}$) at the superdroplet location. The ML model is trained using the superdroplet attributes (e.g. location within grid cell, superdroplet radius, effective supersaturation) as well as the value of filtered Eulerian flow variables at the superdroplet location.  \emph{A priori} analysis of DNS data shows that there is a large difference between $S_{eff}$ and $\filt{S}$ at the superdroplet location, indicating that the well-mixed assumption can lead to large errors in prediction of superdroplet growth rate. On the other hand the trained ML model is able to predict $S_{eff}$, and therefore superdroplet growth rate, with high accuracy. In this work, we will describe the methodology for mapping the actual droplets in DNS to superdroplets, as well as the workflow for training the ML model using DNS data. The accuracy of the ML model will then be assessed for different droplet radii, time instances and superdroplet multiplicity for a given LES grid cell width.  

The rest of text is organized in as follows; Methodology used for generating data from DNS for making superdroplets, proprieties of this data and details of ML model training is provided in the next section. Section \ref{sec:results} details the results and conclusions are presented in section \ref{sec:conclusions}.

\section{ Data and Methodology}\label{sec:methodology}

\subsection{Description of DNS data and droplet evolution}

The data for training the machine learning model is generated by carrying out Direct Numerical Simulations (DNS) for entrainment and mixing in cloud turbulence using the flow solver described in \cite{kumar2013cloud,kumar2014lagrangian}. 

\begin{figure}[h!]
	\centering
    \includegraphics[width=8cm]{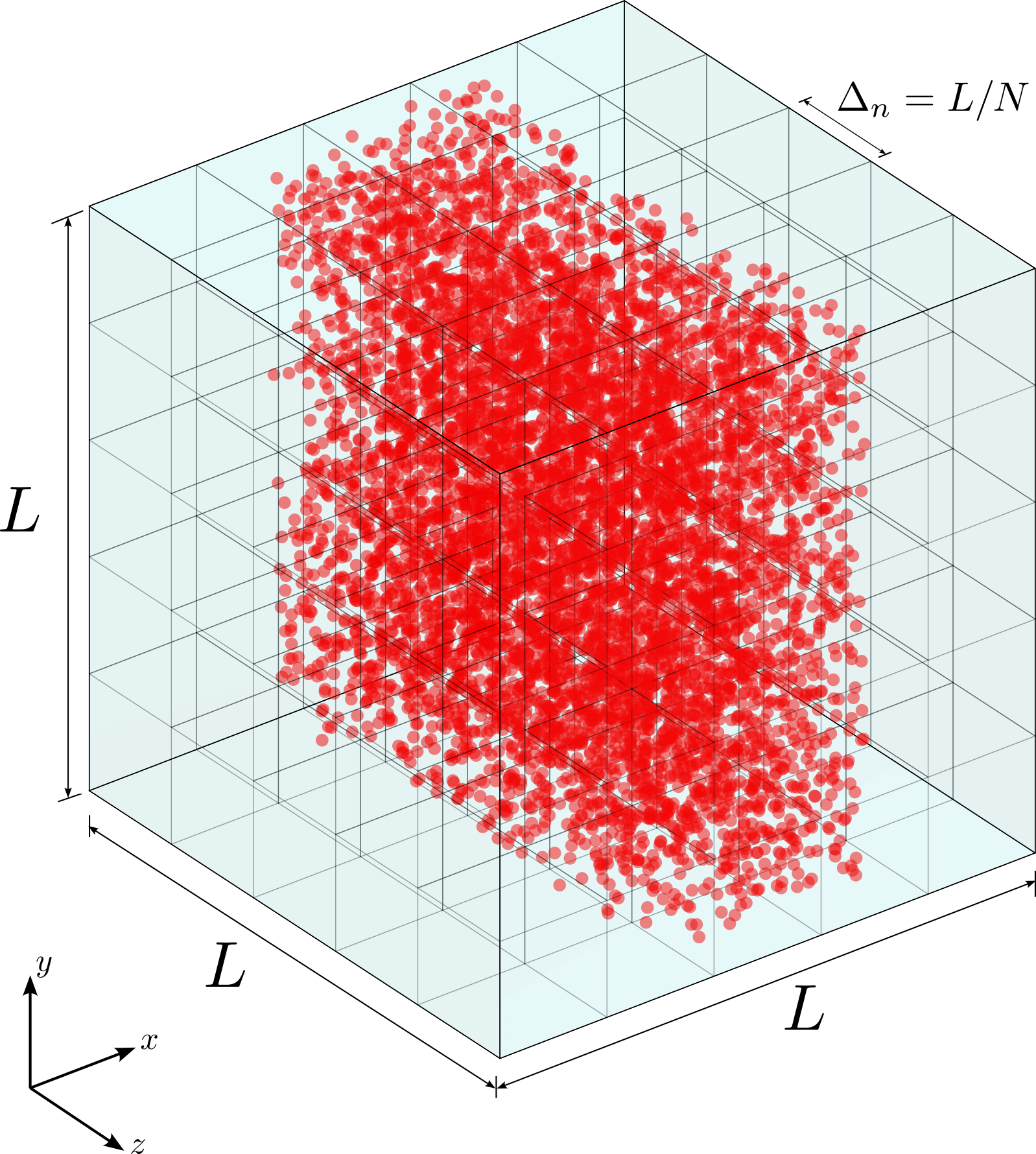}
	\caption{ \small{Schematic showing flow domain with initial configuration of droplets and mesh used for DNS}. }
	\label{fig:domain}
\end{figure}

The computational domain is initialized with a supersaturated slab of vapor containing the cloud droplets over one-third of the computational domain ($x/L\in[1/3,2/3]$)  (Figure \ref{fig:domain}). Droplets with initial radius $15$ microns are randomly placed within the supersaturated vapor slab at time $t=0$, similar to the setup in \cite{kumar2014lagrangian} . A velocity field corresponding to forced homogenous isotropic turbulence with specified average dissipation rate is used to mix the vapor and droplets. The relevant parameters used in the DNS are listed in Table \ref{tab:dnsvalues}. The droplets and vapor field are mixed by the velocity field, and the droplets may condense or evaporate with time, depending on the local supersaturation of the vapor field. The initial temperature field ensures that net initial buoyancy due to temperature and vapor field is zero over the entire domain \cite{kumar2014lagrangian}. 

At any time instant $t$ the DNS data consists of the following fields and quantities, defined over a box of size $L^3$ with $N^3$ Eulerian grid points and $N_d$ droplets as shown in Figure \ref{fig:domain} (see Table \ref{tabsymb} for nomenclature of symbols):
\begin{itemize}
\item Eulerian fields: velocity field $\bu(\bx,t)$, vapor mixing ratio field $q_v(\bx,t)$, temperature field $T(\bx,t)$  collocated equispaced grid points separated by a distance $\Delta_\eta=L/N$, equal to the Kolmogorov length scale in each direction. The supersaturation field can be derived from vapor mixing ratio via the relation $S(\bx,t)=q_v(\bx,t)/q_{v,sat}(\bx,t)-1$
\item Lagrangian fields: Position of droplets $\mathbf{X}^{d,\beta}(t)$, radius of droplet $R_{d,\beta}(t)$, supersaturation of vapor field at droplet location $S_{d,\beta}(t)=S(\mathbf{X}^{d,\beta}(t),t)$, and droplet temperature $T_{d,\beta}(t)=T(\bX^{d,\beta}(t),t)$, with $\beta\in[1,2,..N_d]$ looping over number of droplets $N_d$.
\end{itemize}
Here the subscript/superscript ``$d$" denotes quantities corresponding to the ``actual" droplets. The DNS used in this work has a domain size $L=51.2$ cm, number of grid points $N=512$ along each direction and $N_d=6771519$ actual droplets. The DNS grid cell width is $\Delta_\eta=1$ mm. Importantly, Damkohler number based on the phase relaxation time is $Da=1.95$, indicating that the mixing is  inhomogenous.

\begin{table}[htpb]
	\centering
	\begin{tabular}{|l|l|l|}
        \hline 
        \textbf{Quantity} & \textbf{Symbol} & \textbf{Value} \\
        \hline
        Kinematic viscosity & $\nu$ & $1.5\times 10^{-5} \textrm{ m}^2 \textrm{s}^{-1}$\\
        \hline
        Mean energy dissipation rate & $\avg{\epsilon}$ & $33.75\textrm{ cm}^2 \textrm{s}^{-3}$ \\
        \hline
        Kolmogorov length scale & $\eta_K$ & $1$ mm \\
        \hline
        Kolmogorov time & $\tau_K$ & $0.066$ s\\
        \hline
        Root mean squared velocity & $u_{RMS}$ & $12.5 \textrm{cm}/\textrm{s}$ \\
        \hline
        Large scale turnover time & $T_L$ & $4.1$s \\
        \hline
        Initial droplet radius & $R_0$ & $15$ $\mu$m \\
        \hline
        Initial evaporation time scale & $\tau_{evap}^{0}=-R_0^2/(2 K S_0)$ & $2.1$ s \\
        \hline
        Damkohler number & $\textrm{Da}=\tau_L/\tau_{evap}^0$ & $1.95$ \\
        \hline
        \end{tabular}
        \caption{ \label{tab:dnsvalues} Parameters in DNS simulation}
\end{table}

\begin{table}[htpb]
	\centering
	\begin{tabular}{|l|l|}
 \hline
 \textbf{Symbol} & \textbf{Meaning} \\
\hline 
  $\textbf{x}$, $t$ & Position, time \\
  \hline
  $L$  & Length of the domain \\
  \hline
  $N$  & Number of grid points along each direction \\
  \hline
  $\Delta_\eta$  & DNS grid cell width ($=L/N$) \\
  \hline
   $N_{LES}$ & Number of LES grid cells in each direction \\
  \hline
  $\Delta$ & LES grid cell width ($=L/N_{LES}$) \\
   \hline
    $N_d$ &  Number of actual droplets in flow domain \\
    \hline
    $N_s$ &  Number of superdroplets in flow domain ($\approx N_d/N_m$)\\
    \hline
    $N_m$ & Multiplicity of superdroplets \\
    \hline 
    $N_c$ & Number of contiguous LES grid cells in each direction \\
          &  centered around superdroplet  \\
    \hline 
	$\mathbf{u}(\bx,t)$ & Actual velocity field \\
	\hline
	$q_v(\bx,t)$  & Actual vapor mixing ratio field \\
  \hline
     $q_{v,sat}(\bx,t)$ & Saturated mixing ratio field\\
    \hline
     $S(\bx,t)$ & Actual supersaturation field $(=q_v(\bx,t)/q_{v,sat}(\bx,t)-1)$ \\
     \hline
	 $T(\bx,t)$  &  Actual temperature field \\
	\hline
	$\bX^{d,\alpha}$(t)  & Position of actual droplet with index $\alpha$ \\
	\hline
	$\bX^{s,\lambda}$(t) & Position of superdroplet with index $\lambda$ \\
	\hline
	$R_{d,\alpha}(t)$ &  Radius of actual droplet with index $\alpha$\\
     \hline
     $R_{s,\lambda}(t)$ & Radius of superdroplet with index $\lambda$ \\
	\hline
	$S_{d,\beta}(t)$ & Supersaturation at location of actual droplet with index $\beta$ \\
	\hline
	$T_{d,\beta}(t)$ & Temperature at location of actual droplet with index $\beta$\\
	\hline
	$\filt{\mathbf{u}}(\bx,t)$ &  Filtered velocity field \\
	\hline
	$\filt{q}_v(\bx,t)$ &  Filtered LES vapor mixing ratio \\
	\hline
	$\filt{T}(\bx,t)$ &  Filtered LES temperature \\
     \hline
     $\filt{S}(\bx,t)$ & Filtered supersaturation field \\
    \hline
    $\hat{S}_{s,\lambda}(t)$     &     Filtered  supersaturation of droplet with index $\lambda$ ($=\filt{S}(\bX^{s,\lambda}(t),t)$) \\
	\hline
	$S_{eff,\lambda}(t)$     &     Effective supersaturation of droplet with index $\lambda$\\
  \hline
\end{tabular}
   \caption{\label{tabsymb}: List of symbols}
\end{table}

The droplet radius is evolved via the following equation:
\begin{eqnarray}
\label{eq:drdt}
R_{d,\beta}\frac{d R_{d,\beta}}{dt}&=& K S_{d,\beta}(t)
\end{eqnarray}
where the value of $S_{d,\beta}(t)$ is interpolated from the resolved supersaturation field $S(\bx,t)$ at the droplet location $\bX^{d,\beta}(t)$ and $K$ is a constant which primarily depends on local vapor diffusivity \cite{kumar2014lagrangian}. Simulations based on Large Eddy Simulation (LES) only able to resolve the filtered flow fields, such as $\filt{S}(\bx,t)$, where $\filt{(\cdot)}$ denotes a low-pass spatial filter. A similar coarse-graining is carried out for the droplet dispersion. The superdroplet method  maps clusters of $N_m$ actual droplets to individual superdroplets, so that the entire domain has $N_s\approx N_d/N_m$ actual droplets \cite{shima2009super}. Specifically, a superdroplet with index $\lambda\in\{1,2,..N_s \}$ is associated to a cluster of actual droplets with index belonging to the set $D_\lambda=\{\beta_1,\beta_2,...,\beta_{N_m} \}$ in which $\beta_i\in\{1,2,3,...,N_d\}$. 
The radius $R_{s,\lambda}(t)$ of each superdroplet, located at $\bX^{s,\lambda}(t)$, is evolved via the following equation \cite{shima2009super}:
\begin{eqnarray}
\label{eq:drsdt}
R_{s,\lambda}\frac{d R_{s,\lambda}}{dt}&=& K \hat{S}_{s,\lambda}(t)
\end{eqnarray}
where $\lambda$ denotes the index of the superdroplet and $\hat{S}_{s,\beta}(t)=\filt{S}(\bX^{s,\lambda}(t),t)$ is the value of filtered supersaturation at the superdroplet location. In fact the growth rate of superdroplets should be determined by the growth rate (Eqn. \ref{eq:drdt}) of actual droplets belonging to each superdroplet, which in turn depends on the resolved supersaturation $S$ at the actual droplet location. Before deriving the correct growth rate of superdroplets, we first relate the radius of superdroplets to the associated actual droplets.  Ideally, the superdroplet radius $R_{s,\lambda}$ should be equal to the actual droplets associated with it. However, in a DNS, each superdroplet will be associated with actual droplets belonging to some quantile of droplet radius. In this work, we constrain the mass of superdroplet to be the same as the mass of associated actual droplets, so that for a superdroplet with index $\lambda$:
\begin{eqnarray}
\label{eq:masscon}
R_{s,\lambda}^3(t)&=&\frac{1}{N_m}\sum_{\beta\in D_\lambda} R_{d,\beta}^3(t)
\end{eqnarray} 
Ensuring Eqn. \ref{eq:masscon} implies the following relationship between $\dot{R}_{s,\lambda}$ and the growth rate of associated actual droplets $\{ \dot{R}_{d,\beta};\beta\in D_\lambda\}$:
\bea
\label{eq:drsdt_1}
R_{s,\lambda}\frac{d R_{s,\lambda}}{dt}&=& \frac{1}{R_{s,\lambda} N_m}\sum_{\beta\in D_\lambda} R_{d,\beta}^2\frac{d R_{d,\beta}}{dt}
\eea
Now using Eqns. \ref{eq:drsdt_1} and \ref{eq:drdt}, we can observe that the exact growth rate of a superdroplet with index $\lambda$ satisfies:
\bea
\label{eq:drsdt_correct}
R_{s,\lambda}\frac{d R_{s,\lambda}}{dt}&=& K S_{eff,\lambda}(t) 
\eea
where the \emph{effective} supersaturation $S_{eff,\lambda}(t)$ of the superdroplet is given by:
\bea
\label{eq:seff_def}
S_{eff,\lambda}(t)&=&\frac{1}{R_{s,\lambda} N_m} \sum_{\beta\in D_\lambda} R_{d,\beta} S_{d,\beta}(t)
\eea
Comparing Eqns. \ref{eq:drsdt} and \ref{eq:drsdt_correct}, it is clear that the superdroplet growth rate $S_{s,\lambda}$ obtained using $\tilde{S}$ will in general not agree with the superdroplet growth rate based on $S_{eff}$. If $R_{s,\lambda}\approx R_{d,\beta}$ for $\beta\in D_\lambda$ (which will be true if the radius quantile width $\Delta R$ for actual droplets belonging to $D_\lambda$ is small compared to $R_{s,\lambda}$), then the effective superdroplet supersaturation $S_{eff,\lambda}(t)\approx (1/N_m) \sum_{\beta\in D_\lambda} S_{d,\beta}(t)$ can be interpreted as the average supersaturation at locations of associated actual droplets.  For a given number density of actual droplets, the error between $\hat{S}_{s,\lambda}$ and $S_{eff,\lambda}$ can be particularly large if $N_m$ is small, since $S_{eff,\lambda}$ will represent the average of $S$ over a small volume compared to the volume $\Delta^3$ represented by the filter width. This work will therefore focus on predicting $S_{eff,\lambda}$ for each superdroplet, given filtered flow variables in the vicinity of the superdroplets, as well as superdroplet attributes.

\subsection{Generation of data from DNS}

The data for training, validating and testing the ML model is created from original DNS data that contains detailed information about individual superdroplets along with the filtered flow field in the vicinity of each superdroplet. Each superdroplet contains $N_m$ actual droplets; $N_m$ is typically denoted as the superdroplet multiplicity. The total number of superdroplets in the flow domain is $N_s$ ($\approx N_d/N_{m}$). For each superdroplet, the following information is generated in the training data:

\begin{itemize}
    \item \textbf{Value of filtered Eulerian fields in neighboring grid cells:} Filtered DNS fields $\filt{\bu}$, $\filt{S}$, $\filt{T}$ in the \(N_c \times N_c \times N_c\) contiguous grid cells surrounding the superdroplet. 

    \item \textbf{Lagrangian superdroplet Properties:} The specific attributes of the superdroplet, such as its location $\mathbf{X}^{s,\lambda}(t)$, radius $R_{s,\lambda}(t)$ , and effective supersaturation, $S_{eff,\lambda}(t)$, where $\lambda\in\{1,2,..,N_s\}$ is the index for the superdroplet.   
\end{itemize}

The properties of superdroplets and their surrounding filtered data have been selected as inputs to the model while the output of the model is the effective supersaturation $S_{eff}$ of the superdroplet. Details on generation of data for filtered Eulerian fields and superdroplet properties from DNS data have been given below.

\subsubsection*{(a) Filtered Eulerian fields}

\begin{figure}[h!]
\centering
\subfigure[]{\label{dns}
\includegraphics[scale=0.4]{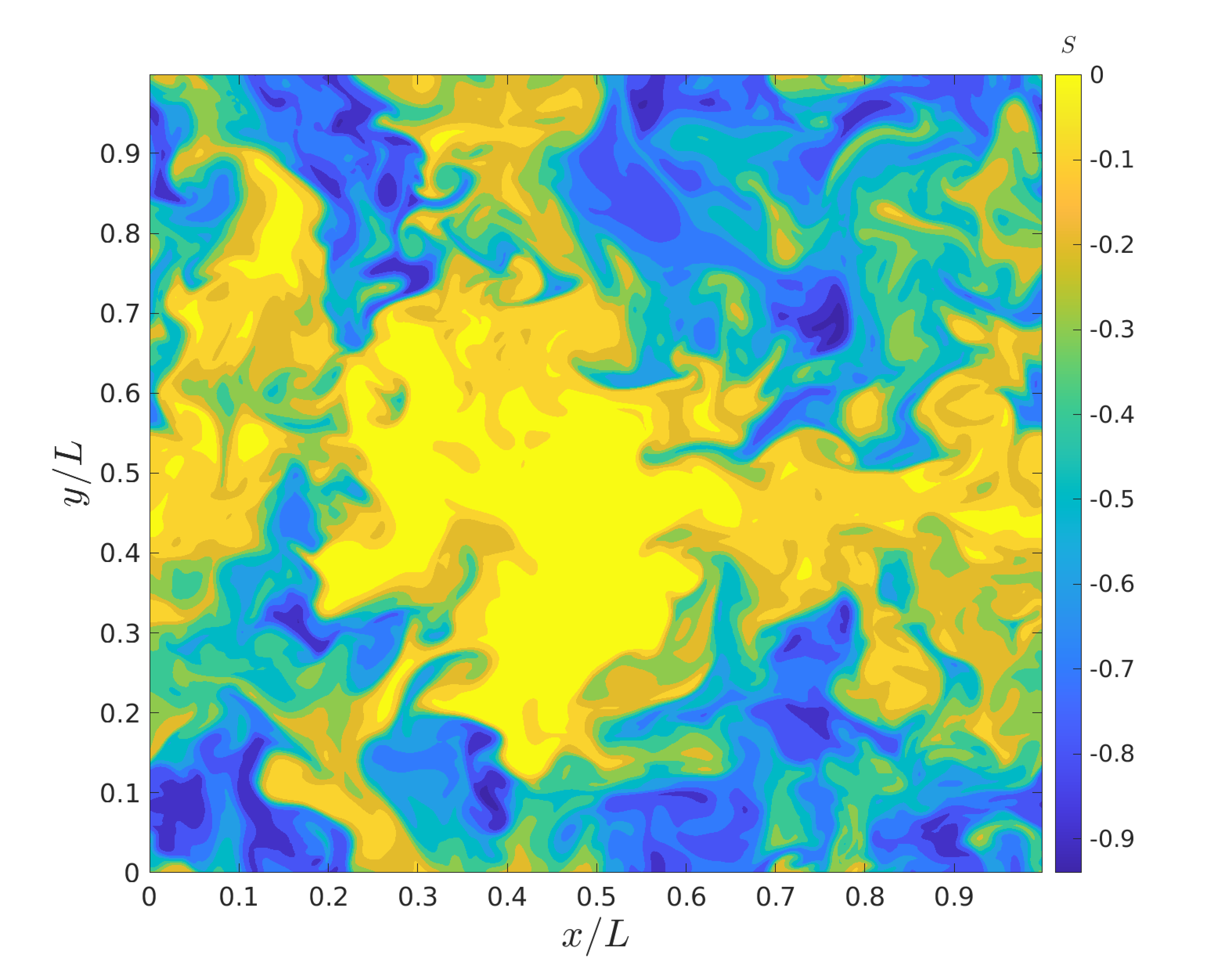}}
\subfigure[]{\label{les}
\includegraphics[scale=0.4]{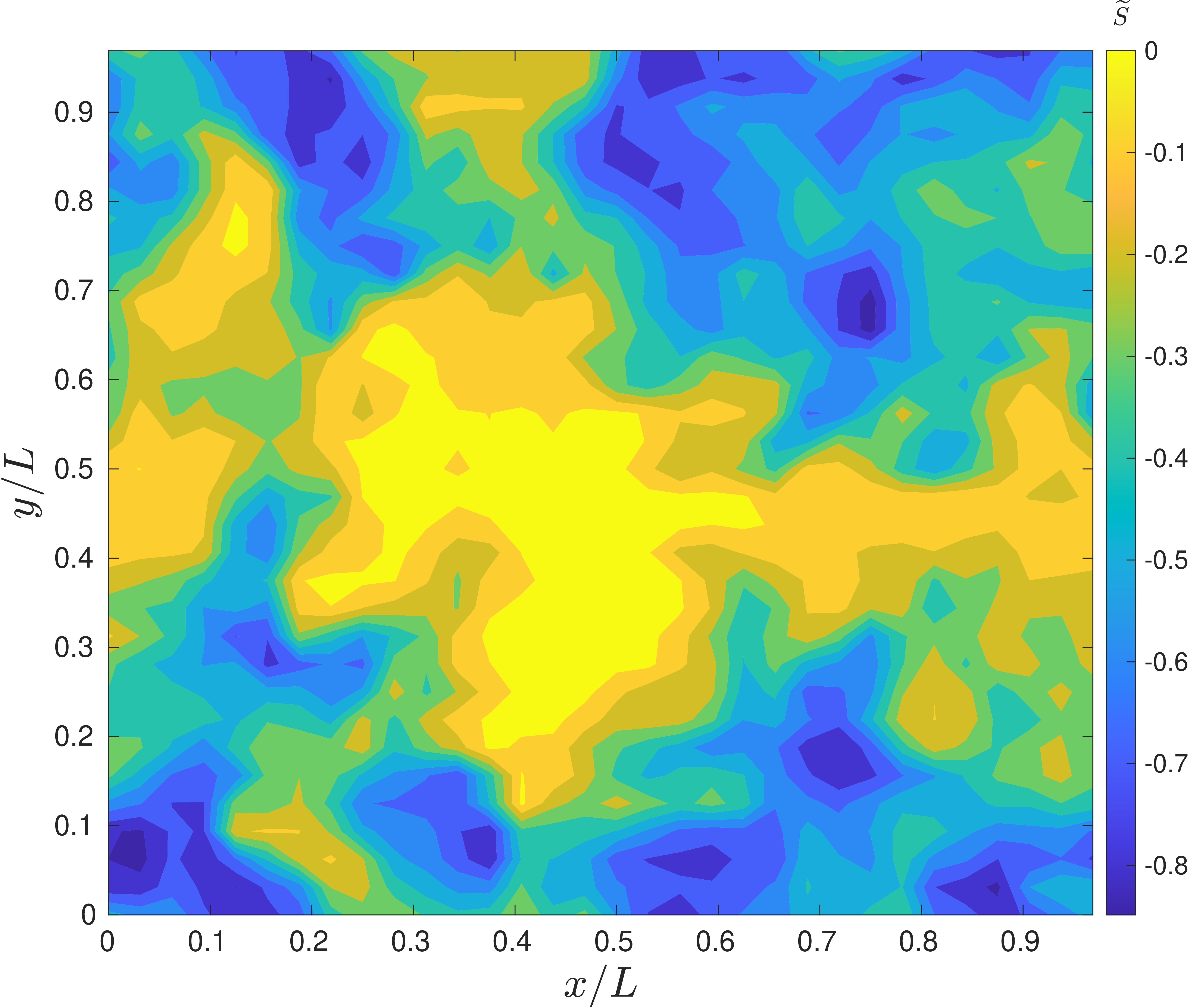}}
\caption{ \small{ Filled isocontours at $t=2.5$ seconds (5000 time steps) and $z=\Delta/2$ for (a) Supersaturation field $S(\bx,t)$ from DNS data and (b) filtered supersaturation field $\tilde{S}(\bx,t)$. Here $\Delta=L/32$ for the box filter $\tilde{(\cdot)}$. } }
\label{fig:dns_les}
\end{figure}

 The filtered Eulerian fields ($\filt{\bu}$, $\filt{S}$, $\filt{T}$) are generated by applying a box filter with width $\Delta=L/N_{LES}$ on the DNS Eulerian fields. Here, $N_{LES}$ is the number of grid cells along each direction in the filtered DNS field. 
In this work, we specifically train the model for $\Delta=16 \Delta_\eta$ ($=16$ mm), which corresponds to $N_{LES}=32$.  
 The filtered LES fields are denoted as $\tilde{u}_i(\bx,t)$, $\tilde{q}_v(\bx,t)$, $\tilde{T}(\bx,t)$, $\filt{S}(\bx,t)$ which are collocated over the $N_{LES}^3$ filtered DNS grid cells. The isocontours for $S(\bx,t)$ and $\filt{S}(\bx,t)$ at $z=\Delta/2$ and $t=5$ s have been shown in  Figure~\ref{fig:dns_les}. Clearly, $\filt{S}$ captures the large scale features of the DNS supersaturation field $S$, while filtering out the small scale features. 
 
\subsubsection*{(b) Superdroplet Data}

\begin{figure}[h!]
	\centering
	\includegraphics[scale=0.75]{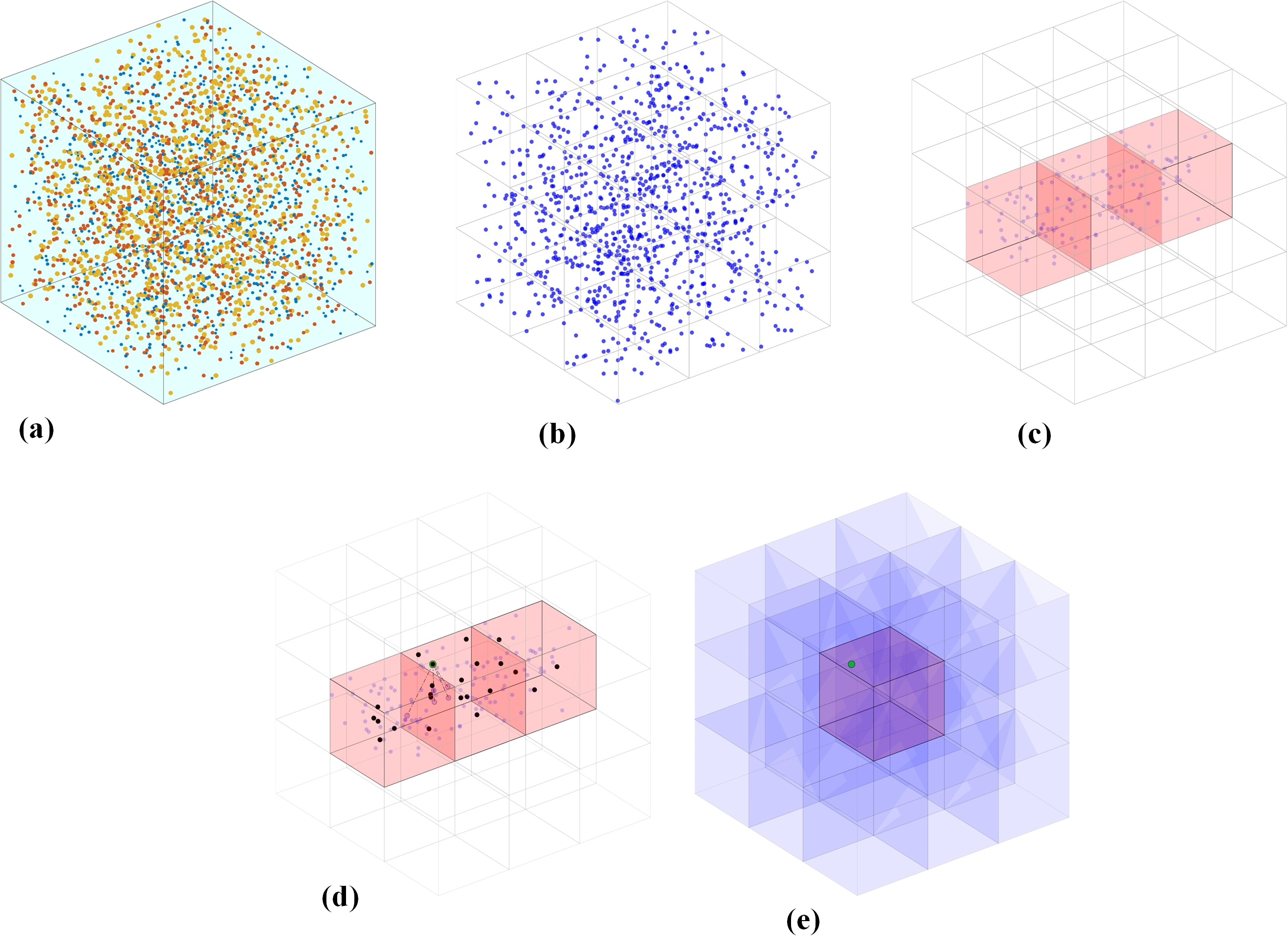}
	\caption{\small{Illustration of superdroplet generation process in DNS domain. (a) Separation of the droplets into quantiles based on their radius. Dots of a specific color represents a different quantile. (b) Actual droplets belonging to a radius quantile overlaid on LES grid (a 3×3×3 LES grid is shown here for illustration). (c) Pencil of LES cells along $x$. (d) Aggregation of droplets (small dots) into superdroplets (large dots). (e) A single superdroplet (green dot) has been shown along with $N_c\times N_c\times N_c$ contiguous neighboring LES grid cells (blue shading, with $N_c=3$. The purple LES grid cell contains the superdroplet. }}
	\label{fig:superprocess}
\end{figure}

There are two major stages during the generation of superdroplet data. The first stage consists of generation of superdroplet location $\{\bX^{s,\lambda}(t);\lambda\in\{1,2,..,N_s\} \}$ from the location of actual droplets $\{\bX^{d,\beta}(t);\beta\in\{1,2,..,N_d\} \}$. 
 The various steps for generating superdroplet location has been illustrated in Figure~\ref{fig:superprocess}. At any given time $t$ the $N_d$ actual droplets are partitioned into $N_q=10$ quantiles based on their radius (Figure~\ref{fig:superprocess} (a)), in which each quantile consists of $\approx N_d/N_q$ actual droplets. The actual droplets belonging to a particular radius quantile (Figure~\ref{fig:superprocess}(b)), are then partitioned into smaller sets of actual droplets belonging to pencils of LES grid cells aligned along $x$ (Figure~\ref{fig:superprocess}(c)). For $N_d^p$ actual droplets located within each pencil, the position of $N_s^p \approx N_d^p/N_m$ superdroplets are generated over region occupied by the pencil, using the following procedure.
 \begin{itemize}
  \item  A cumulative distribution function (CDF) $p=F(x)$ is first constructed for the $x$ location of the subset of actual droplets within each pencil.
  \vspace{-2 mm}
  \item  Random uniformly distributed values for $p$ between 0 and 1 are then generated and mapped to corresponding values $x=F^{-1}(p)$ to yield the $x$ coordinate of the superdroplets. 
    \vspace{-2 mm}
  \item  The $y$ and $z$ coordinates of the superdroplet are randomly generated within each pencil from a uniform distribution; here, we ignore the variation of droplet concentration within each LES grid cell along the $y$ and $z$ direction, since the pencil is oriented along $x$. 
    \vspace{-2 mm}
  \item  Once the spatial positions of the superdroplets  have been determined for all the pencils (Figure~\ref{fig:superprocess}(d)), a \( K \)-nearest neighbors (KNN) search algorithm is employed to identify \( N_m \) actual droplets that constitute each superdroplet belonging to the radius quantile. 
    \vspace{-2 mm}
  \item  At the end of this step, we obtain the set of actual droplet indices $D_\lambda$ for every superdroplet index $\lambda$. 
 \end{itemize}
 
\begin{figure}[h!]
	\centering
	\subfigure[]{\label{fig:adrops}
		\includegraphics[scale=0.4]{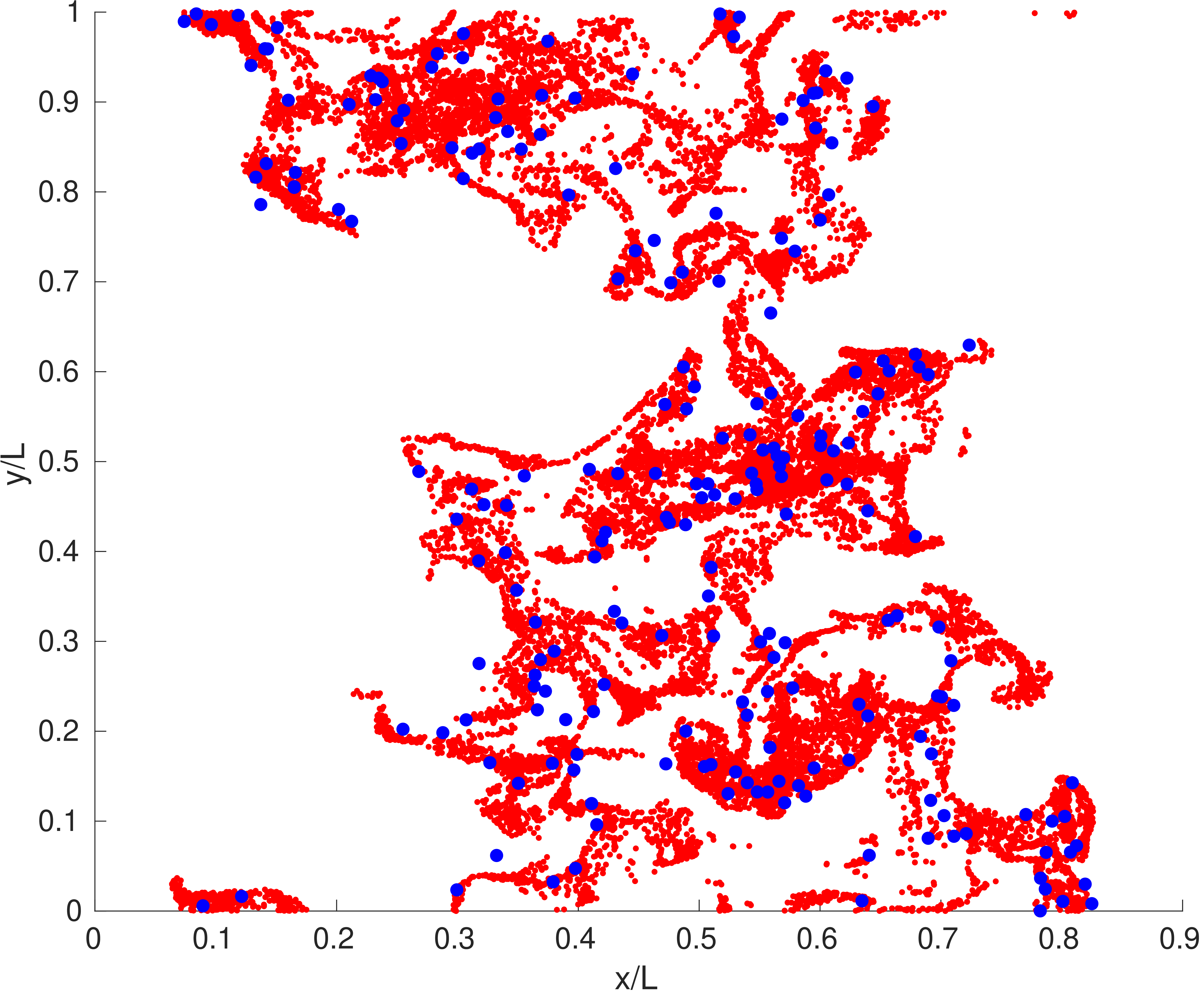}}
	\subfigure[]{\label{fig:sdrops}
		\includegraphics[width=0.55\textwidth, height=7 cm]{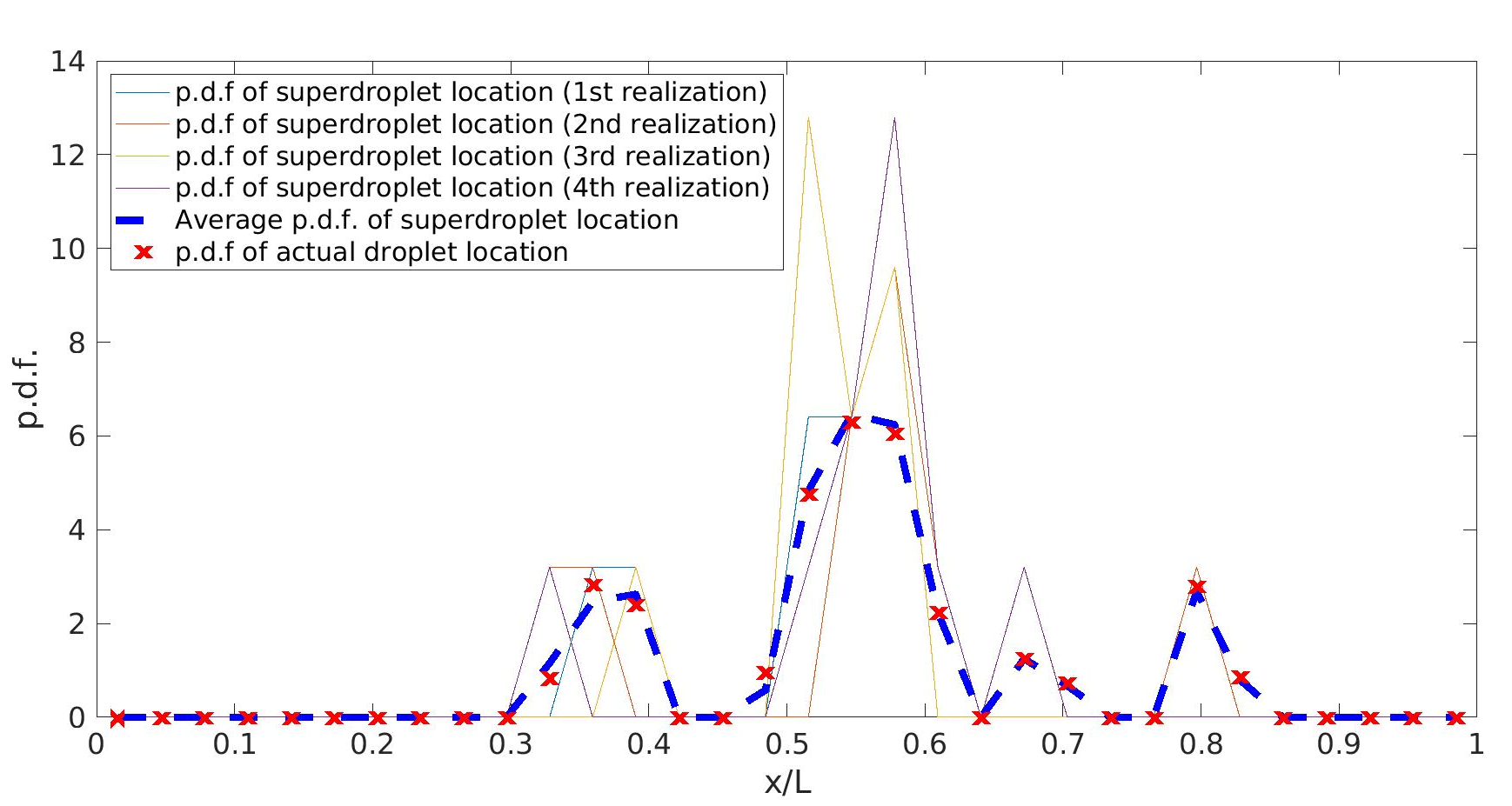}}
	\caption{\small{Lagrangian data at $t=1$ second, $z/\Delta\in[4,\,5]$ ($\Delta=L/32$) and droplet radius quantile $R\in[15.09,\,15.1]\mu m$ showing (a) Location of superdroplets (\tikzcircle[blue, fill=blue]{3pt}) superimposed on location of actual droplets (\tikzcircle[red, fill=red]{1.5pt}) projected on the $x-y$ plane.  (b) Probability distribution function (p.d.f) for $x$ location of 4 different realizations of superdroplets (\textcolor{green}{\full}), superimposed over p.d.f. for $x$ location of actual droplets (\textcolor{red}{$\times$}) and p.d.f. for $x$ location of superdroplets averaged over 100 realizations (\textcolor{blue}{$\dashed$}). Here all droplets and superdroplets belong to the  pencil $y/\Delta\in[4,\,5]$, $z/\Delta \in[4,\,5]$}}.
	\label{fig:pdfeqv}
\end{figure}

 This process is repeated for  each radius quantile. Eqn. \ref{eq:masscon} is used to calculate the effective radius $R_{s,\lambda}(t)$ of each superdroplet from the radius of the associated actual droplets, and Eqn. \ref{eq:seff_def} is used to calculate the effective supersaturation $S_{eff,\lambda}$ of the superdroplet.  Figure~\ref{fig:pdfeqv}(a) shows the location of actual droplets and superdroplets superimposed on each other at $t=1$ second, $z/\Delta\in[4,5]$ and $R_d\in[15.09,\,15.1]\mu\,m$. The distribution of actual droplets show fine, filament-like features. On the other hand, the distribution of superdroplets captures the large-scale spatial features of the distribution of actual droplets. This is not surprising, since the number concentration of superdroplets is much lesser compared to the concentration of actual droplets. Figure~\ref{fig:pdfeqv}(b) shows the probability distribution (p.d.f.) of the $x$ location of actual droplets superimposed over several realizations of the p.d.f. of $x$ location of superdroplets within a pencil, as well as the average of 100 realizations of p.d.f. of $x$ location of superdroplets within the pencil. Clearly, a single realization of superdroplet distribution does not capture the p.d.f. of the actual droplets accurately, but the average p.d.f. of superdroplets from 100 realizations agrees quite well with the p.d.f. of the actual droplets.

\subsubsection*{(c) Structure of data}

For any given time instant, the DNS data used to train, validate and test the ML algorithm is organized into a list of records, where each record corresponds to the properties of a superdroplet and the filtered Eulerian field in its neighborhood. Each record contains the following data for superdroplet with index $\lambda$:
\begin{enumerate}
    \item \textbf{Line 1:} Lists  superdroplet attributes (4 features), including offset location from the floor of its LES grid cell $\delta \mathbf{X}_{s,\lambda}(t)$, and droplet radius $R_{s,\lambda}$. The value of effective supersaturation \( S_{eff,\lambda}(t) \), serves as the label for prediction. 
    \item \textbf{Lines 2 to \( \mathbf{N_c^3 + 1 }\):} Each line here corresponds to one of the $N_c\times N_c\times N_c$ neighboring LES grid cells, and lists 5 filtered DNS features for the grid cells ($\filt{S}$, $\filt{T}$ and all 3 components of $\filt{\bu}$)
   \end{enumerate}

Thus, the number of features per superdroplet in the training data is \( 4+5N_c^3  \). For \( N_c = 3 \) (i.e., $3\times 3\times 3$ surrounding cells) used in our work, the dataset for each superdroplet consists of 139 features.
A single dataset is generated for a specific \emph{time-instant} of the simulation, fixed multiplicity of superdroplet ($N_m$) and fixed filter width $\Delta$.

Further pre-processing is required before this data can be used for training. As part of pre-processing, the data is loaded and parsed from multiple text files, each representing a different time step in the cloud simulation generated in the previous section. The features are then extracted and then normalized to ensure consistency across all variables, which is crucial for optimal performance in many machine learning algorithms.

The final step of pre-processing involves creating a comprehensive feature set by concatenating the scaled superdroplet properties with LES cell data into a 1D feature vector. Each 1D feature vector is then paired with a label representing the effective supersaturation around each superdroplet, which acts as the target variable for prediction. These feature vectors and their corresponding labels are subsequently used for training, validation, and testing of the machine learning model.

\subsection{Training, validation and testing of ML Model}
\begin{figure}[h!]
    \centering
    \includegraphics[width=14cm]  {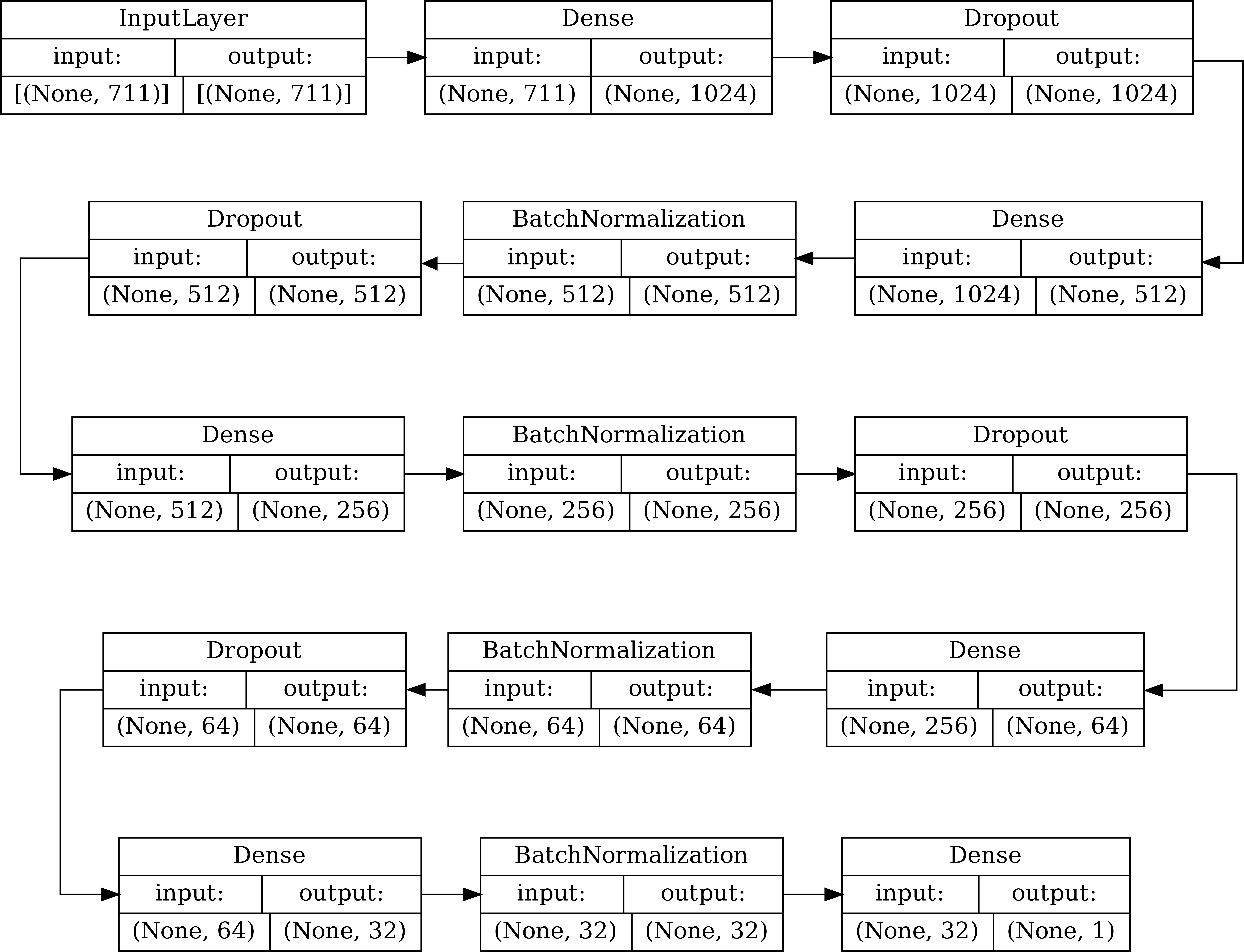}
    \caption{\small{A schematic of the ML model developed in this study.}}
    \label{fig:mlmodel}
\end{figure}

The data is split into training, validation, and testing sets, with $70\%$ used for training and $15\%$ each for validation and testing. An MLP model is built using TensorFlow's Keras API. The architecture consists of a 1024-unit input layer followed by four hidden layers with 512, 256, 64, and 32 units, respectively. Each hidden layer includes batch normalization and a $30\%$ dropout rate to prevent overfitting. The output layer is a single unit with linear activation, appropriate for regression tasks. The model is compiled using the Adam optimizer with a learning rate of 0.0005 and Mean Squared Error (MSE) as the loss function. The model is trained for 100 epochs with a batch size of 256, and its performance is validated using the validation set. A schematic of the model is shown in Figure~\ref{fig:mlmodel}. 

\begin{figure}[h!]
	\centering
	\includegraphics[width=0.5\linewidth]{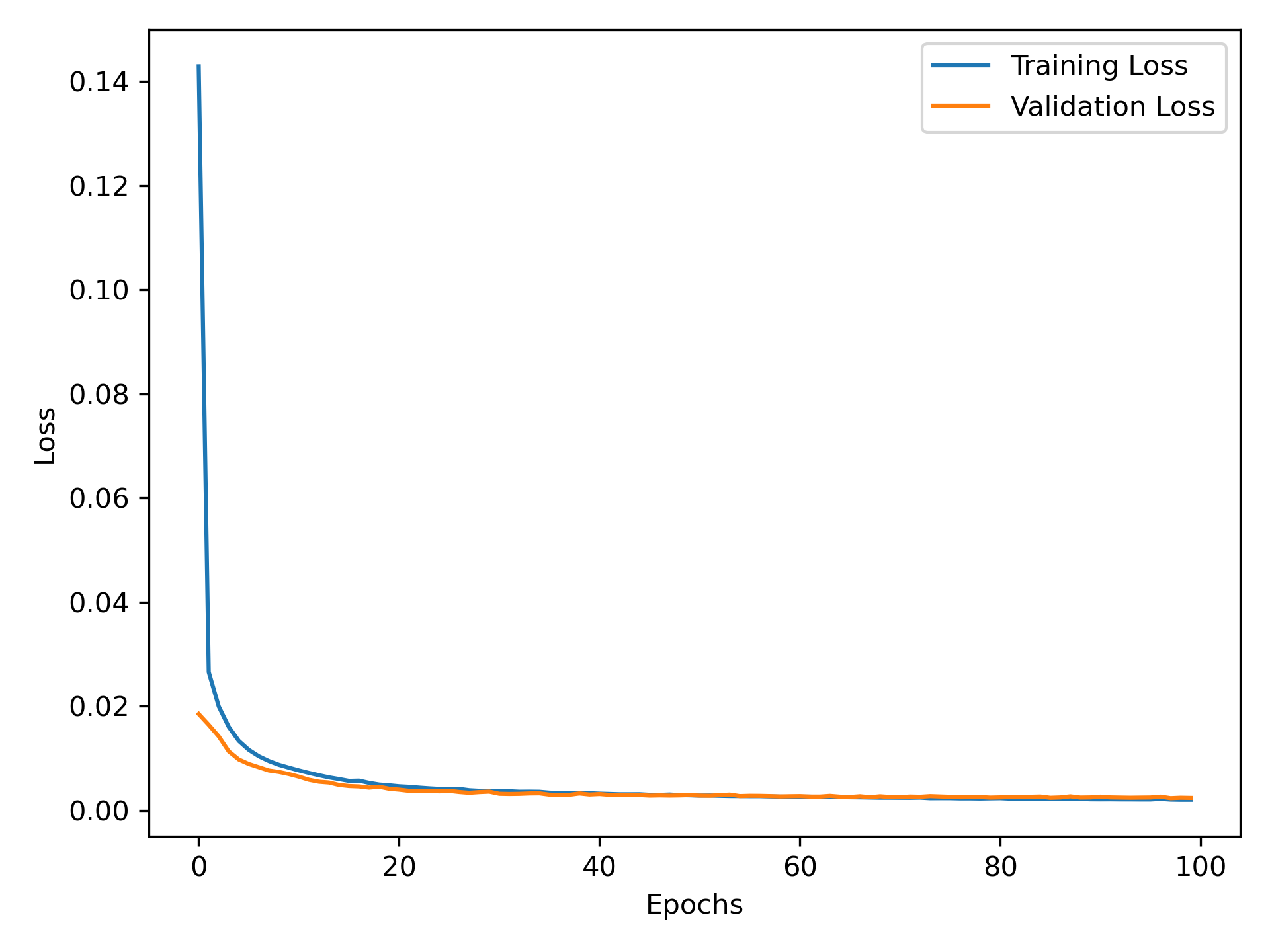}
	\caption{\small{Evolution of training and validation losses for dataset with $N_m=300$.}}
	\label{fig:tvloss}
\end{figure}

\begin{figure}[h!]
    \centering
    \includegraphics[width=0.6\linewidth]{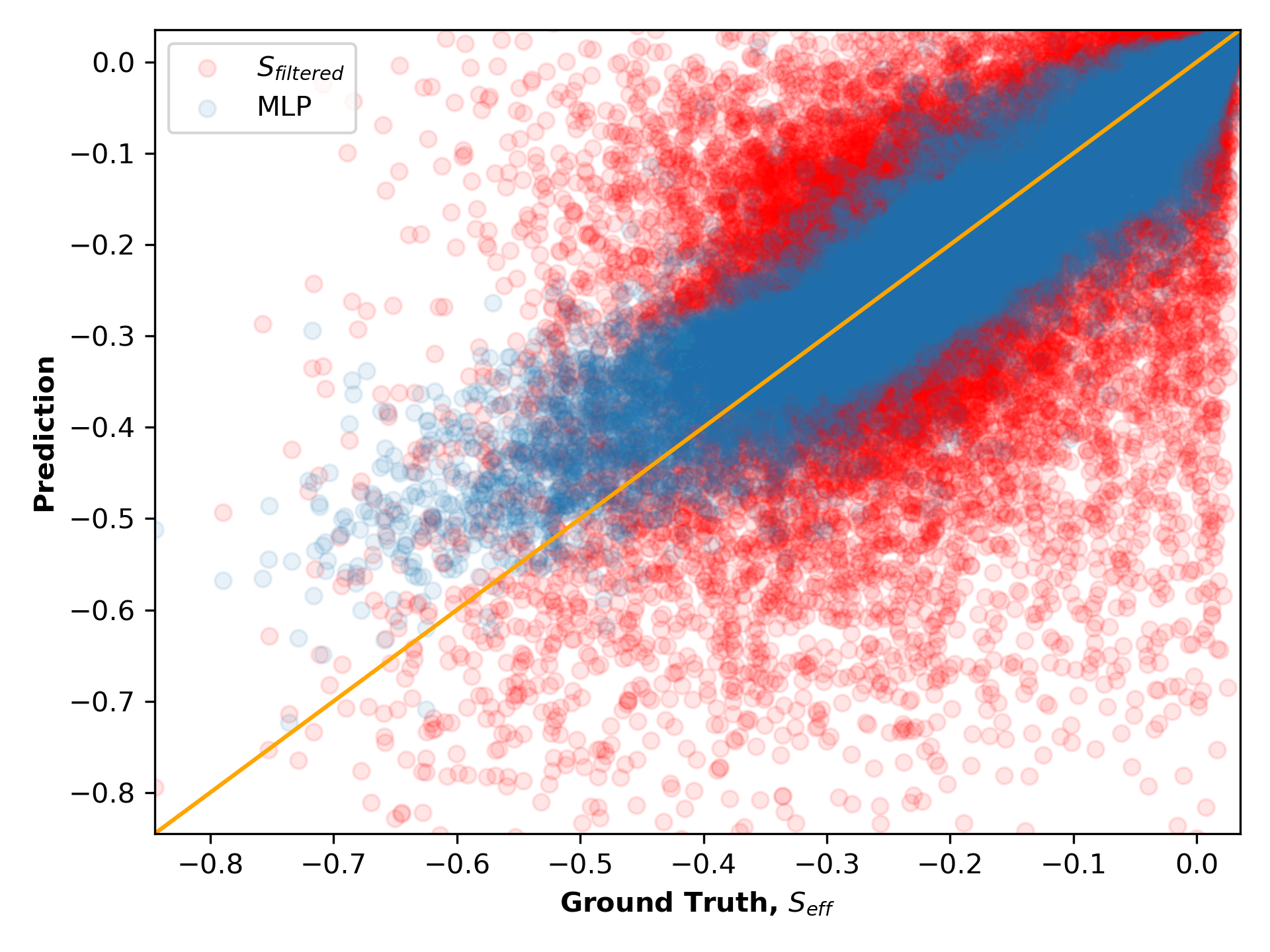}
    \caption{\small{Scatter plot with $S_{eff,\lambda}$ on abscissa and effective supersaturation predicted by the ML model (MLP) and well mixed model (\(S_{\text{filtered}}\equiv \filt{S}(\bX^{s,\lambda}(t),t)\)) in the coordinate. This plot corresponds to data set with $N_m=300$.}}
    \label{fig:r2plot}
\end{figure}

The ML model is trained on three datasets, each corresponding to a multiplicity of $N_m=$100, 200, and 300, respectively. The value of filter width $\Delta=L/32$ is fixed. The loss plot for one such dataset is presented in Figure~\ref{fig:tvloss} showing that, for the data set with $N_m=300$, 100 epochs are enough for training the ML model. Each dataset corresponds to 10 time instances in the DNS, ranging from $t=0.5$ to $t=5$ seconds. The metric used for evaluating the output of the ML model (i.e. predicted supersaturation $S_{pred,\lambda}$ for superdroplet with index $\lambda$) is the R² score based on the difference between $S_{eff,\lambda}$ and $S_{pred,\lambda}$.

\section{Results}\label{sec:results}

The result for the dataset with a multiplicity of $N_m=300$ are discussed from here onwards. 
The scatter plot in Figure ~\ref{fig:r2plot} compares the supersaturation (\(S_{\text{pred},\lambda}(t)\)) predicted by the machine learning (ML) model and the filtered supersaturation at the superdroplet’s location (\(S_{\text{filtered},\lambda}(t)\equiv \filt{S}(\bX^{s,\lambda}(t),t)\)) against the effective supersaturation (\(S_{\text{eff},\lambda}(t)\)), which serves as the ground truth, as well as the label for the ML model. The data has been plotted for all the 10 time instants. The $S_{eff}-S_{filtered}$ scatter plot (red points) show large scatter around the principal diagonal in the graph, with $R^2=0.27$, indicating large differences between the two quantities at superdroplet locations. The $S_{eff}-S_{pred}$ scatter plot (blue points) shows far less error, with $R^2=0.89$, indicating fairly accurate predictions of effective supersaturation by the ML model, and a sinnificant improvement with respect to the well-mixed assumption.

\begin{figure}[h!]
	\centering
	\includegraphics[width=0.8\linewidth]{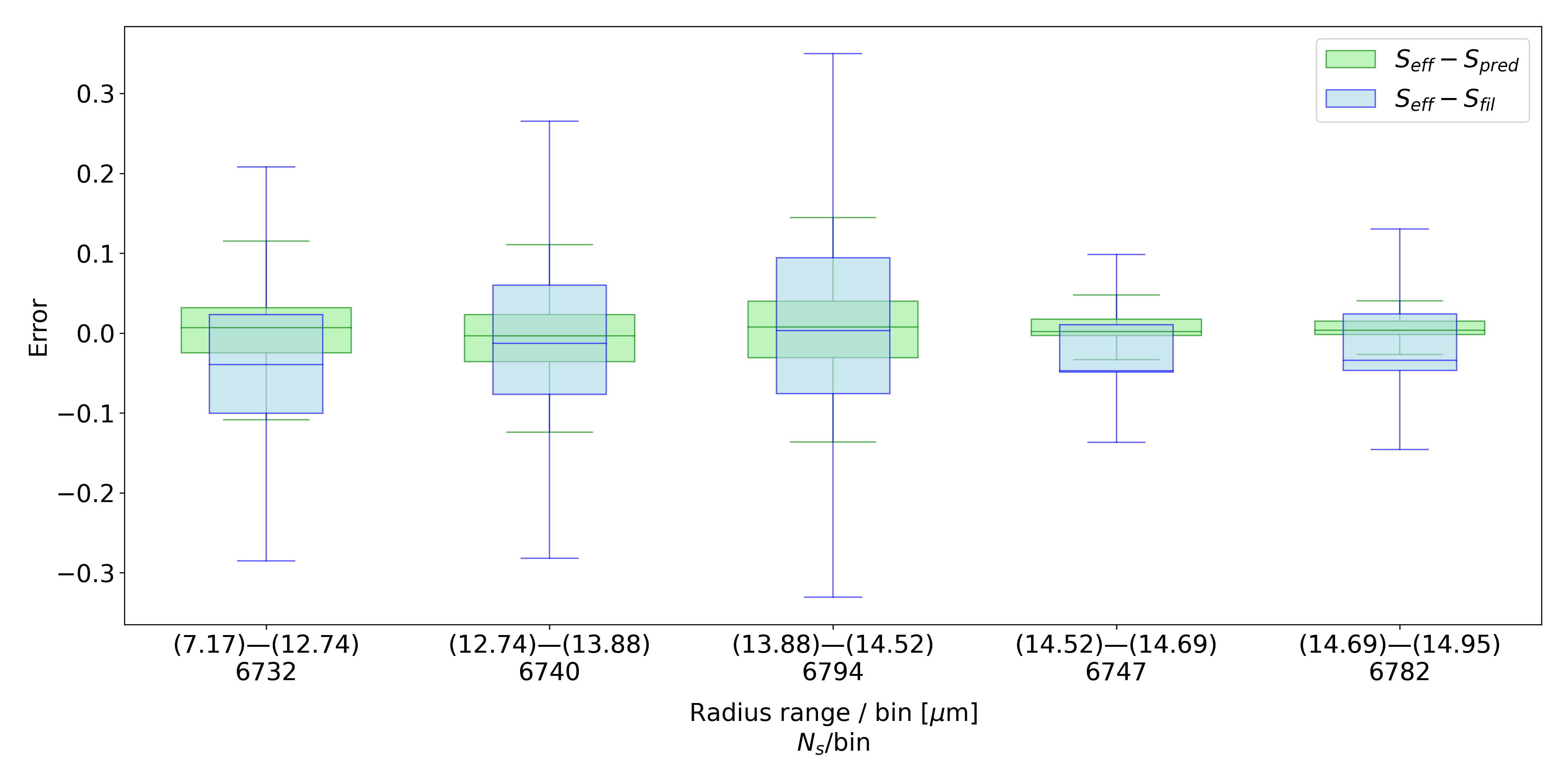}
	\caption{\small{Boxplots of prediction errors for different superdroplet size quantiles, comparing the ML model's supersaturation predictions (\(S_{\text{eff}} - S_{\text{pred}}\)) with filtered DNS data (\(S_{\text{eff}} - S_{\text{filtered}}\)).} }
	\label{fig:qtileserr}
\end{figure}

The error distribution in supersaturation predictions across different quantiles of superdroplet sizes is illustrated in Figure~\ref{fig:qtileserr}. The figure presents boxplots showing the $L_2$ norm error between (\(S_{\text{eff}}\)) and $S_{\text{filtered}}$ and two models: the machine learning (ML) model (blue) and the well-mixed assumption (green). The error is calculated over all superdroplets across 10 different time instances. 
The results clearly demonstrate that the ML model achieves significantly lower prediction errors for effective supersaturation, with errors typically falling within a narrow range of $\pm 0.15$. In contrast, the well-mixed assumption yields a broader error range, between $\pm 0.35$. Interestingly, both models show the largest prediction errors for medium-sized superdroplets ($R_s \in (13.88, 14.52),\mu m$), while the error is notably smaller for the largest droplets ($R_s \in (14.69, 14.95),\mu m$).

This trend can be explained by the fact that the largest droplets are generally found in more homogeneously mixed regions of the cloud, where predictions are easier and more accurate. On the other hand, medium-sized droplets are often located near the edges of supersaturated regions, where inhomogeneous mixing occurs, making accurate predictions more challenging.  Overall, the ML model significantly reduces the prediction error for effective supersaturation compared to the well-mixed assumption, across the full range of droplet sizes.

\begin{figure}[h!]
	\centering
	\includegraphics[width=0.8\linewidth]{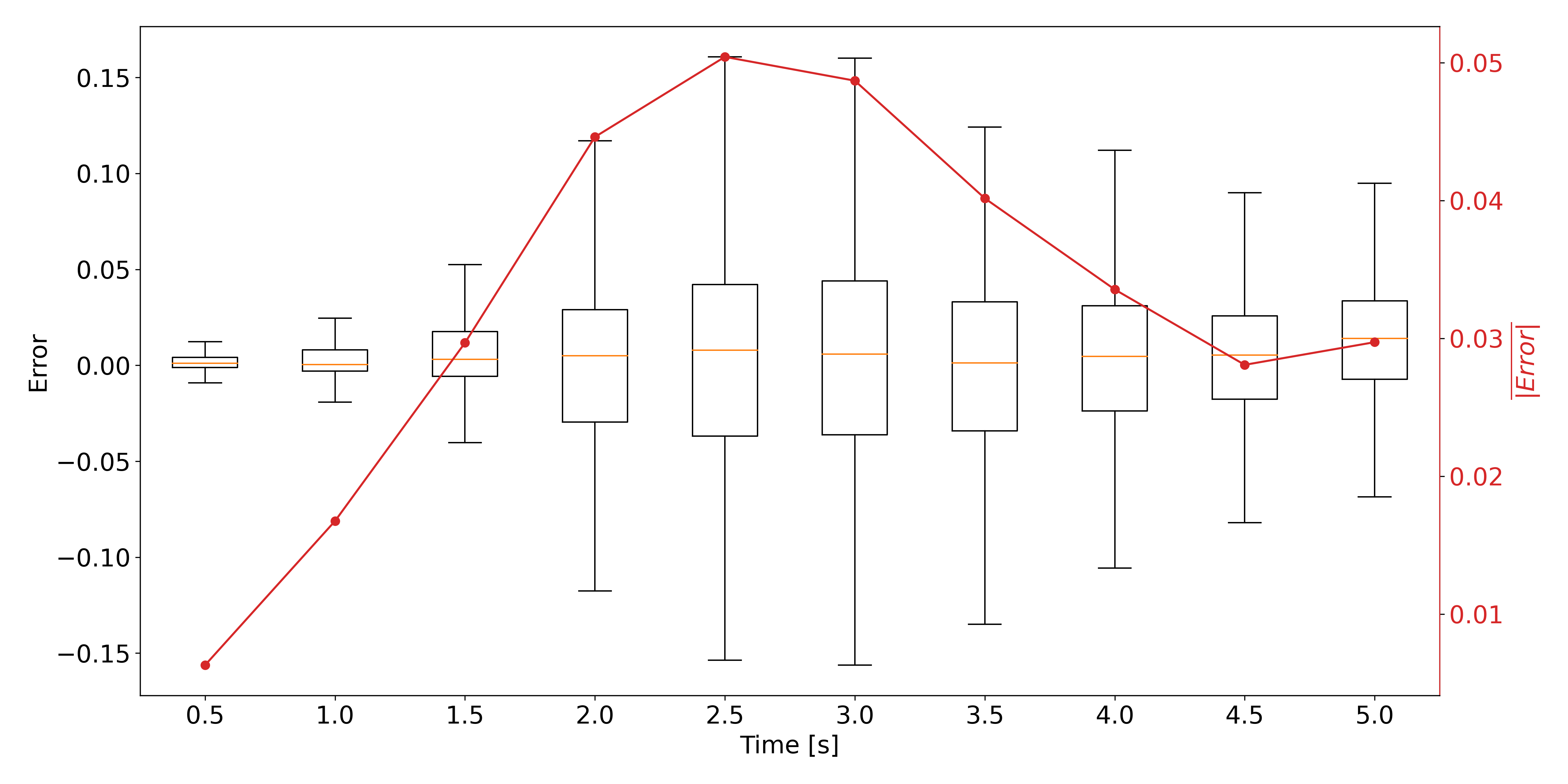}
	\caption{\small{Box plot of prediction error effective supersaturation by ML model with respect to time superimposed on graph showing mean of the absolute prediction error with respect to time.} }
	\label{fig:enter-label}
\end{figure}

As the simulation progresses, the prediction error evolves in tandem with the changing dynamics of the system, as illustrated in Figure~\ref{fig:enter-label}. In the initial stages of the simulation, when the majority of superdroplets are located in a well-mixed, supersaturated environment, the prediction errors are relatively low. During this phase, the cloud's vapor and droplet distribution are more uniform, which allows both the machine learning (ML) model and the well-mixed assumption to make accurate predictions of effective supersaturation. The homogeneity of the environment minimizes variability in the input features, leading to reduced errors across the board.

However, as the simulation advances, complex physical processes such as droplet evaporation, condensation, and turbulent mixing intensify. These processes introduce greater variability in the distribution of vapor and droplets within the cloud. The interaction between regions of varying humidity, supersaturation, and droplet sizes becomes more intricate, leading to more heterogeneous conditions. As a result, the prediction errors begin to rise, especially during intermediate time steps. At this point, the range of prediction errors expands, and the mean absolute error reaches its peak. 


Towards the later stages of the simulation, the vapor and droplet fields gradually return to a more uniform and well-mixed state. This transition reduces the complexity of the physical interactions within the cloud, allowing the models to regain accuracy in their predictions. As the level of inhomogeneous mixing decreases, the prediction errors correspondingly diminish. By the end of the simulation, the vapor and droplets once again exhibit a more balanced distribution, and the prediction errors stabilize at a lower level, similar to the initial phase of the simulation. Thus, the evolution of prediction error over time directly mirrors the cloud's transition from a well-mixed to a more turbulent and complex state, before settling back into homogeneity. 



\begin{figure}[ht]
	\centering
	\includegraphics[width=0.7\linewidth, height=0.31\textheight]{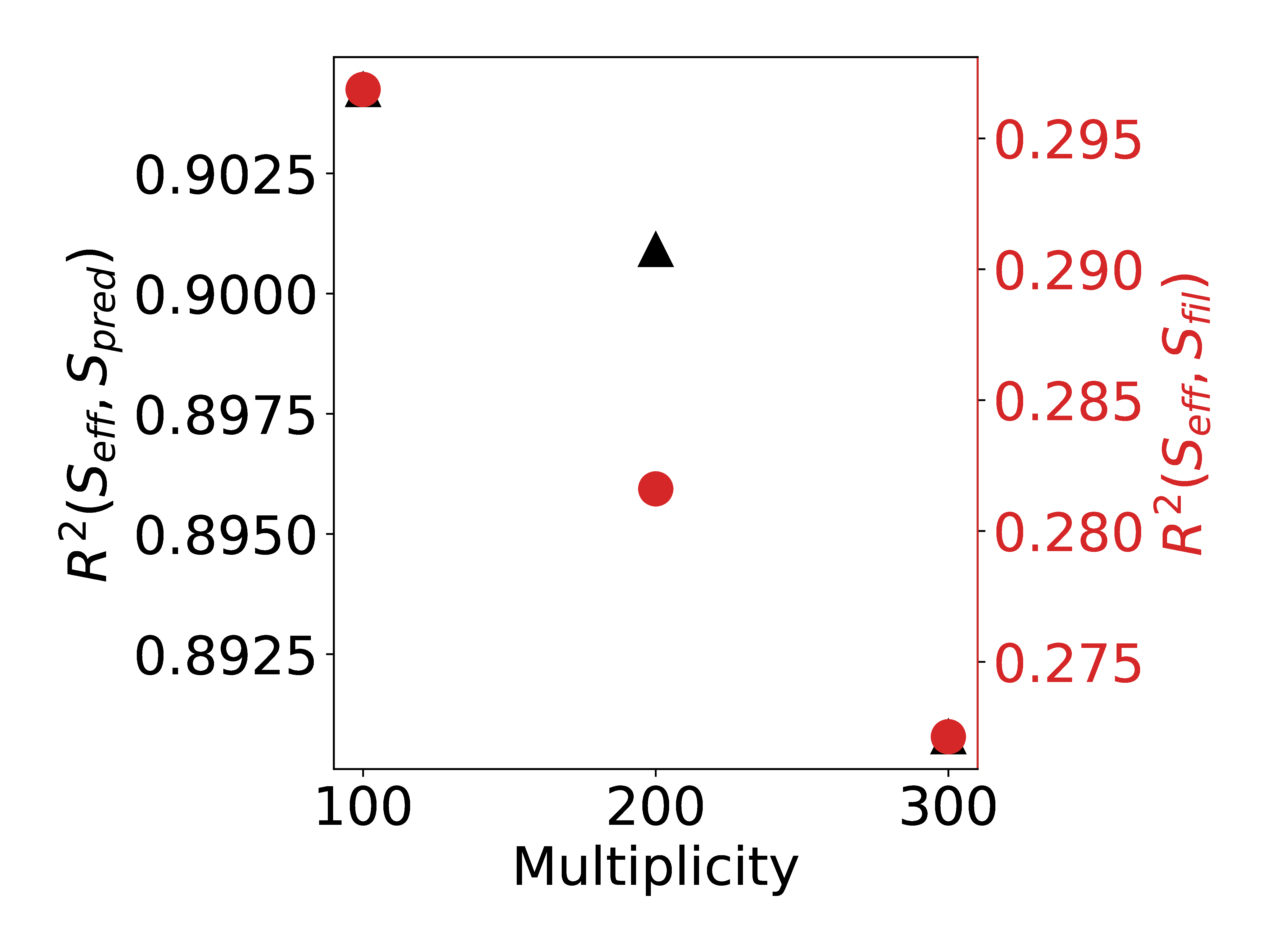}
	\caption{\small{Comparison of prediction accuracy ($R^2$) at different multiplicities of superdroplets and $\Delta=L/32$.}}
	\label{fig:nmcomp}
\end{figure}

When comparing the $R^2$ values between the ML model and the well-mixed assumption at different multiplicities ($N_m$), there is minimal variation, with changes in $R^2$ staying below 3\% for both models (Figure~\ref{fig:nmcomp}). However, a slight decrease in $R^2$ occurs with higher multiplicities, likely due to the reduced number of superdroplets ($N_s$) available for training as $N_m$ increases. This reduction in dataset size introduces greater variance in the prediction of effective supersaturation, slightly affecting the performance of both models.

Overall, the ML model consistently reduces prediction errors in effective supersaturation compared to the well-mixed assumption across a range of droplet sizes and time steps. The evolution of errors aligns with the physical processes in the simulation, such as mixing and evaporation, while the analysis of $R^2$ values emphasizes the impact of dataset size on prediction accuracy, particularly at higher multiplicities.

\section{Conclusions}\label{sec:conclusions}

This work introduced a technique to map actual droplets from DNS data to superdroplets, corresponding to filtered DNS data while preserving mass consistency. The effective supersaturation of each superdroplet is calculated by ensuring that the mass attributed to it matches the total mass of the actual droplets it represents. Our analysis reveals a significant discrepancy between the filtered and effective supersaturation, indicating that the well-mixed assumption used in the LES solves introduces substantial errors when predicting the growth rates of superdroplet radii.

We developed a machine learning (ML) model trained on filtered DNS data and superdroplet properties to address this issue. The results demonstrate that the ML model effectively predicts the effective supersaturation, achieving high accuracy (e.g., $R^2 = 0.89$ for $N_m = 300$). This performance is consistent across a range of multiplicities ($N_m$) and filter widths ($\Delta = L/32$), significantly improving predictions compared to the well-mixed assumption.

The evolution of prediction errors throughout the simulation highlights the robustness of the ML model in capturing complex cloud dynamics, especially during periods of intense mixing and phase changes. The model's ability to handle the variability in droplet environments suggests its potential for broader applications. In future work, we aim to extend this approach by incorporating the ML model into LES to model the evolution of superdroplet size spectra during turbulent mixing at cloud edges. The proposed approach will enhance our understanding of cloud microphysical processes and provide more accurate predictions in real-world scenarios involving inhomogeneous mixing.


 \bibliographystyle{basic} 
\bibliography{ref_ML_droplet_growth}

\end{document}